\documentclass[conference]{IEEEtran}
\IEEEoverridecommandlockouts
\usepackage{mathtools}
\usepackage{bm}
\usepackage{graphicx}
\usepackage{booktabs}
\usepackage{cite}
\usepackage{url}
\usepackage{microtype}
\usepackage{siunitx}
\usepackage{amsmath,amsfonts,amssymb,amsthm, bm}
\newtheorem{theorem}{Theorem}

\newtheorem{corollary}{Corollary}
\newtheorem{definition}{Definition}
\newtheorem{remark}{Remark}
\newtheorem{assumption}{Assumption}
\usepackage{tikz}
\usetikzlibrary{arrows.meta,positioning,fit,calc,shapes.geometric,shapes.multipart}
\usetikzlibrary{positioning,calc,arrows.meta}
\usetikzlibrary{positioning,arrows.meta,fit,backgrounds,decorations.pathreplacing}
\tikzset{
  arrow/.style={-Latex, line width=0.8pt},
  block/.style={draw, rounded corners=2pt, align=center, minimum height=6mm, inner sep=2pt, font=\small},
  op/.style={block, fill=gray!10},
  var/.style={block, fill=blue!5},
  gate/.style={block, fill=orange!12},
  delay/.style={block, fill=yellow!15},
  legendbox/.style={draw, rounded corners=2pt, inner sep=2pt, font=\scriptsize, fill=white},
  lbl/.style={font=\scriptsize, inner sep=1pt}
}

\expandafter\def\expandafter\normalsize\expandafter{%
    \normalsize%
    \setlength\abovedisplayskip{0pt}%
    \setlength\belowdisplayskip{0pt}%
    \setlength\abovedisplayshortskip{-2.5pt}%
    \setlength\belowdisplayshortskip{0pt}%
}

\title{Network-Optimised Spiking Neural Network (NOS) Scheduling for 6G O-RAN: Spectral Margin and Delay-Tail Control}

\author{\IEEEauthorblockN{1\textsuperscript{st} Muhammad Bilal}
\IEEEauthorblockA{\textit{School of Computing and Communications} \\
\textit{Lancaster University}\\
Lancaster, United Kingdom \\
m.bilal@ieee.org}
\and
\IEEEauthorblockN{2\textsuperscript{nd} Xiaolong Xu}
\IEEEauthorblockA{\textit{school of computer and software} \\
\textit{Nanjing University of Information Science and Technology}\\
Nanjing, China \\
xlxu@ieee.org }

}

\begin{document}
\maketitle

\begin{abstract}
This work presents Network-Optimised Spiking Neural (NOS) based delay-aware scheduler for 6G radio access. The proposed scheme couples a bounded two-state kernel to a clique-feasible, proportional-fair (PF) grant head, where the excitability state serves as a finite-buffer proxy and a recovery state acts as a supresser for repeated grants, while neighbour pressure is injected along the interference graph through delayed spikes. A small-signal analysis yields a delay-dependent threshold \(k_\star(\Delta)\) and a spectral margin \(\delta=k_\star(\Delta)-gH\rho(W)\) that compresses topology, controller gain, and delay into a single design parameter. Under light assumptions on arrivals, we prove geometric ergodicity for \(\delta>0\) and obtain sub-Gaussian backlog and delay tail bounds with exponents proportional to \(\delta\). We present a numerical study, aligned with the analysis and a DU compute budget, compares NOS scheduler with proportional fair (PF) and delayed backpressure (BP) baselines across interference topologies with a 5–20~ms delay sweep. With a single gain fixed at the worst spectral radius, NOS scheduler sustains higher utilisation and smaller P99.9 delay while remaining clique-feasible on integer PRBs. 
\end{abstract}


\section{Introduction}
Schedulers in cellular radio access must translate fast fading, interference constraints, and compute limits into timely integer PRB grants. Two families of radio access schedulers dominate practice and theory: 1) Backpressure-type policies are throughput-optimal in multihop networks and provide a rigorous stability template, but their delay behaviour can be fragile under observation noise and actuation latency in fast timescales \cite{TassiulasEphremides1992,Bonald2003MobiCom}, 2) Proportional fair (PF) scheduling delivers a robust throughput–fairness compromise and remains the workhorse in LTE/NR, with practical integer mapping and long-term averaging \cite{Capozzi2013PFsurvey}. In either of these two, the dependence on near–real-time control delays is typically not treated as a first-class design variable, even though it is central in disaggregated O-RAN, where xApps tune policies over tens of milliseconds and DUs execute per-slot logic \cite{Polese2023ORAN,Bonati2023OpenRANGym,Wu2024R3,Longhi2025TailORAN}.

We propose the NOS scheduler, which is a spiking scheduler that preserves networking semantics and exposes a single spectral-margin parameter $\delta$ that aggregates delay, interference topology, and control strength. The NOS scheduler's design builds on the general NOS model presented in \cite{bilal2025NOS}. In this design, each bearer carries a bounded excitability that represents finite buffers and a recovery credit that discourages repeated grants; neighbour pressure is injected through delayed spikes weighted by the interference graph. This modelling choice is inspired by compact spiking systems that combine rich dynamics with low computational cost \cite{bilal2025NOS,Izhikevich2004}, without requiring neuromorphic hardware for deployment \cite{Liu2024EEDSNN}.

In the NOS scheduler, a local linearisation produces a delay-aware threshold \(k_\star(\Delta)\) and a spectral margin \(\delta=k_\star(\Delta)-gH\rho(W)\). The margin serves three roles. First, it yields a stability certificate akin to backpressure, but with delay entering as a formal design variable. Second, it links topology through \(\rho(W)\) and controller gain \(g\) to a single policy knob. Third, it controls tail exponents for backlog and delay, which in turn track performance at fixed radio cap. This view complements PF’s long-term fairness and connects to classical opportunistic scheduling \cite{Liu2001Opportunistic}. We place the NOS scheduler in an O-RAN setting in which the near-RT RIC adjusts the coupling (g) and thresholds at tens-of-milliseconds cadence, while the O-DU executes the 1 ms per-slot loop with clique-feasibility checks on integer PRBs. This split maps DU compute budgets and observation lags directly into the effective delay, which then feeds the spectral margin \(\delta \) \cite{Polese2023ORAN,Bonati2023OpenRANGym}. Compared with delayed backpressure variants and DRL-based controllers, NOS provides an explainable baseline with a single safety-relevant parameter that can be audited before integration, and it does not require per-slot E2 control.

In this paper we makes four contributions. First, we introduce delay-aware spiking scheduler (NOS) that couples a bounded two-state kernel to a clique-feasible, PF-compatible spike-to-PRB map on integer PRBs. Second, we present a small-signal analysis, which yields a delay-aware threshold \(k_\star(\Delta)\) and a spectral margin \(\delta=k_\star(\Delta)-gH\rho(W)\). For \(\delta>0\) we prove geometric ergodicity and derive sub-Gaussian backlog and delay bounds. Third, we present a reproducible \(\delta\)-proxy study that folds DU compute and observation lag into the effective delay and compares NOS with PF and delayed backpressure on pair2, line4, and ring8, using a single coupling \(g\) calibrated once by tail parity at the worst spectral radius and \(\Delta=20\) ms. Finally, we analyse NOS in O-RAN: the DU runs the per-slot loop with integer feasibility checks, while the near-RT RIC tunes \(g\) and thresholds on a tens-of-milliseconds cadence.



\newcommand{\LeftY}{-14mm}

\begin{figure*}[h!]
\centering
\begin{tikzpicture}[
  font=\fontsize{7pt}{7pt}\selectfont,
  node distance=5mm and 10mm, >=LaTeX,
  box/.style={draw, rounded corners=1.5pt, align=center, minimum height=6mm,
              inner xsep=1.8mm, inner ysep=1.2mm, fill=white},
  ue/.style={draw, rounded corners=1.5pt, minimum width=7mm, align=center,
             inner xsep=1mm, inner ysep=0.9mm, fill=white},
  arrow/.style={-{Latex[length=1.8mm,width=1.1mm]}, line width=0.5pt},
  dashedarrow/.style={-{Latex[length=1.8mm,width=1.1mm]}, line width=0.5pt, dashed},
  groupbox/.style={draw, rounded corners=2.5pt, densely dotted, line width=0.7pt, fill opacity=0.06},
  gLeft/.style   ={groupbox, draw=blue!60,   fill=blue!40},
  gNOS/.style    ={groupbox, draw=teal!70!black,  fill=teal!50},
  gSpike/.style  ={groupbox, draw=orange!80!black, fill=orange!60},
  gCtrl/.style   ={groupbox, draw=purple!70!black, fill=purple!50}
]

\node[box] (nos) at (0,0)
{\textit{NOS kernel}\\[-0.2mm]
\scriptsize bounded $f_{\text{sat}}(v)$; recovery $u$;\\
\scriptsize leak $-\chi\,(v-v_{\mathrm{ref}})$; \\
neighbour $g\,W\,S(t-\Delta)$;\\
\scriptsize arrival drive $\eta(t)$};

\node[box, below=5mm of nos, text width=3.5cm, align=left] (stab)
{\textbf{Stability \& tails}\\
\scriptsize margin $\delta>0$ ensures geom.\ ergodicity;\\
\scriptsize delay envelope $k_\star(\Delta)$;\\
\scriptsize tail scale $\propto \delta(\mu_{\min}\bar x)^2$};

\node[box, below=5.5mm of stab, text width=4.3cm, align=center] (timeline)
{$\Delta_{\mathrm{eff}}=\Delta_{\mathrm{sig}}+\Delta_{\mathrm{obs}}+\Delta_{\mathrm{DU}}$\\
\scriptsize near-RT updates: thresholds and gain $g$};

\node[box, left=22mm of nos, yshift=\LeftY, text width=3.2cm] (arr)
{Arrivals\\\scriptsize compound-Poisson; finite moments};

\node[ue, above=2.8mm of arr.north, xshift=-11mm] (ue1) {UE$_1$};
\node[ue, right=1.2mm of ue1] (ue2) {UE$_2$};
\node[ue, right=1.2mm of ue2] (ue3) {UE$_3$};
\node[ue, right=1.2mm of ue3] (uen) {UE$_N$};
\draw[dashed, gray!60] (ue1) -- (ue2) -- (ue3) -- (uen);

\node[box, below=4mm of arr, text width=3.2cm] (queue)
{Slot queues\\\scriptsize $s_i=\mu_i x_i$;\; per-slot $q_i$ update\\
\scriptsize Units: $q_i=\varsigma v_i$};

\path let \p1=(arr.north west), \p2=(arr.north east) in
  coordinate (aN1) at ($(\p1)!0.15!(\p2)$)
  coordinate (aN2) at ($(\p1)!0.35!(\p2)$)
  coordinate (aN3) at ($(\p1)!0.65!(\p2)$)
  coordinate (aN4) at ($(\p1)!0.85!(\p2)$);
\draw[arrow] (ue1.south) to[out=-90,in=90] (aN1);
\draw[arrow] (ue2.south) to[out=-90,in=90] (aN2);
\draw[arrow] (ue3.south) to[out=-90,in=90] (aN3);
\draw[arrow] (uen.south) to[out=-90,in=90] (aN4);

\draw[arrow]
  (queue.east) .. controls ($(queue.east)+(8mm,0)$) and ($(nos.west)+(-8mm,0)$) ..
  (nos.west);

\node[box, right=10mm of nos, text width=3.5cm] (spike)
{\textbf{Spike} $S_i$\\\scriptsize trigger $\mu v/(\bar r+\varepsilon)\ge v_{\mathrm{th}}$};
\node[box, above=5mm of spike, text width=3.5cm] (reset)
{\scriptsize soft reset: $v\!\leftarrow\!c+(v-c)e^{-\nu\tau_{\mathrm{rst}}}$,\; $u\!\leftarrow\!u{+}d$};
\node[box, below=5mm of spike, text width=3.5cm] (smooth)
{\scriptsize smoothing: $r\!\leftarrow\!(1-\zeta)r+\zeta S$\\
\scriptsize suppression: $\tilde r\!\leftarrow\!r/(1+\sum w\,S(t-\Delta))$};
\node[box, below=5mm of smooth, text width=3.5cm] (pf)
{\scriptsize PF fairness:\; $\bar r$ EWMA,\; $w=\tilde r\,\mu/(\bar r+\varepsilon)$};
\node[box, below=5mm of pf, text width=3.6cm, fill=orange!12] (cont)
{\textbf{Clique allocation}\\\scriptsize $x^{\mathrm{cont}}\!\propto\!\gamma_{\mathcal{C}}\,w$ within clique $\mathcal{C}$};
\node[box, below=3.5mm of cont, text width=3.6cm] (intmap)
{\scriptsize floor→drop→renormalise→\\water-fill};
\node[box, below=3.5mm of intmap, text width=3.6cm, fill=gray!3] (phy)
{\textbf{PHY grants}\\\scriptsize feasible PRBs $\le\gamma_{\mathcal{C}}$};

\draw[arrow] (nos.east) -- (spike.west);
\draw[arrow] (reset.west) to[out=180,in=60] (nos.north);
\draw[arrow] (spike) -- (smooth);
\draw[arrow] (smooth) -- (pf);
\draw[arrow] (pf) -- (cont);
\draw[arrow] (cont) -- (intmap);
\draw[arrow] (intmap) -- (phy);
\draw[arrow] (phy.west) -| ($(pf.west)+(-0.4,0)$) |- (pf.west);

\draw[arrow] (nos.south) -- (stab.north);
\draw[arrow] (stab.south) -- (timeline.north);
\draw[dashedarrow]
  ([xshift=-1.5mm]timeline.west)
   .. controls +(-15mm,15mm) and +(-15mm,-5mm) ..
  ([xshift=-1.5mm]nos.west);

\draw[dashedarrow]
  ([xshift=-0.8mm]spike.west)
  to[out=180,in=25]
  node[pos=0.40, above=1.5pt, xshift=-1pt, fill=white, inner sep=0.5pt, font=\scriptsize]
  {$S_{j\neq i}(t-\Delta)$}
  ([yshift=-0.5mm]nos.north east);

\begin{pgfonlayer}{background}
  \node[gLeft, inner sep=2.5mm, fit=(ue1)(uen)] (uegroup) {};
  \node[gLeft, inner sep=3.5mm, fit=(uegroup)(arr)(queue)] {};
  \node[gNOS,  fit=(nos)(stab)] {};
  \node[gSpike,fit=(spike)(reset)(smooth)(pf)(cont)(intmap)(phy)] {};
  \node[gCtrl, fit=(timeline)] {};
\end{pgfonlayer}

\end{tikzpicture}
\caption{Left group (UEs, arrivals, queues) feeds the NOS kernel via a curved arrow. The right chain performs spike generation, fairness, clique allocation and integer PRB mapping. Dashed arrows denote neighbour spikes ($j\neq i$) and near-RT feedback via $\Delta_{\mathrm{eff}}$.}
\label{fig:nos_overview_2col}
\end{figure*}
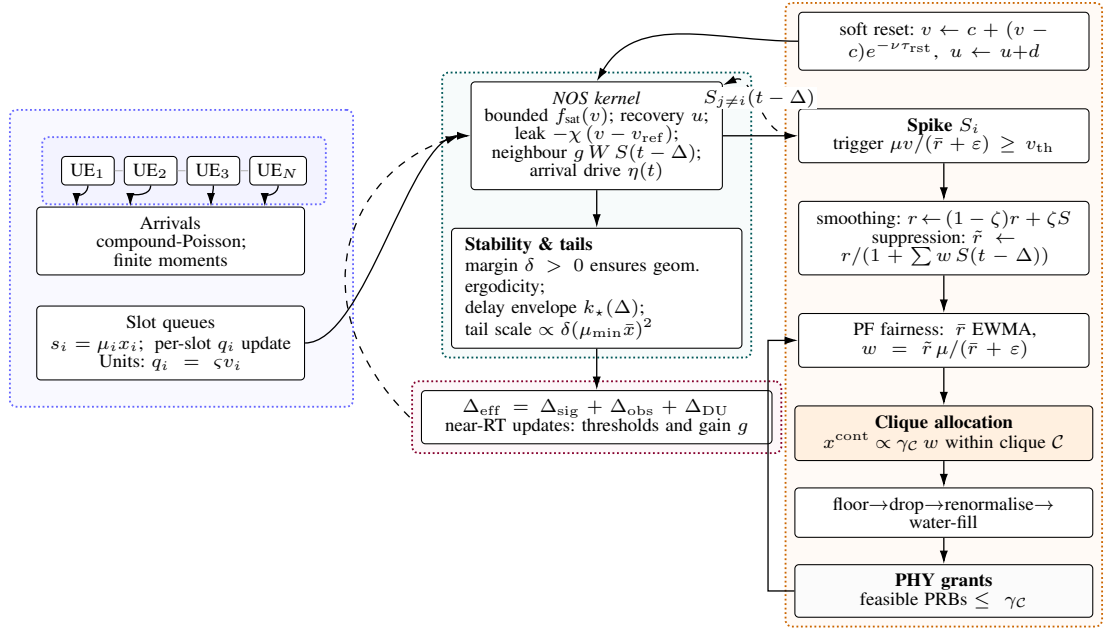
\vspace{-4.5mm}
\section{System Model}
\label{sec:system}
We consider \(N\) bearers indexed by \(i\in\{1,\ldots,N\}\).
In slot \(t\in\mathbb{Z}_{\ge 0}\), bearer \(i\) has backlog \(q_i(t)\), compound-Poisson arrivals \(a_i(t)\) with mean \(\lambda_i\) and finite exponential moments, PRB fraction \(x_i(t)\in[0,1]\), and per-PRB rate \(\mu_i(t)\).
\begin{align}
s_i(t)&=\mu_i(t)\,x_i(t), \label{eq:queue_service}\\
q_i(t+1)&=\max\!\{0,\,q_i(t)+a_i(t)-s_i(t)\}. \label{eq:queue}
\end{align}
We encode Interference by a non-negative matrix \(W=[w_{ij}]\) with spectral radius \(\rho(W)\). Users belong to interference cliques \(\mathcal{C}\) enforcing \(\sum_{i\in\mathcal{C}}x_i(t)\le 1\). We use \(q_i=\varsigma\,v_i\) with \(\varsigma>0\) to set units; \(\varsigma\) fixes the queue scale. Near-RT control signals are delayed by \(\Delta\ge 0\) slots. We target DU-side slot scheduling; the near-RT case uses the same model with \(g\) and thresholds updated every tens of milliseconds.
For analysis we use the NOS state dynamics in continuous time and treat \eqref{eq:queue} as a semantic reference for integer-PRB scheduling and reporting. To align means, choose the noise mean to match arrivals and express the service leak in centred form about an operating-point reference $v_{\mathrm{ref}}$:
\begin{align}
\begin{split}
\mathbb{E}[\eta_i(t)]=\varsigma^{-1}\lambda_i,\;
-\chi\,v_i\;\leadsto\;-\chi\,(v_i-v_{\mathrm{ref}}),\\
\chi\,v_{\mathrm{ref}}=\varsigma^{-1}\,\mathbb{E}[\mu_i(t)\,x_i(t)]  
\end{split}
\end{align}
Then $\mathbb{E}[\dot v_i]=\varsigma^{-1}\big(\lambda_i-\mathbb{E}[\mu_i x_i]\big)$ and $q_i=\varsigma [v_i]_+$ agree in mean at the chosen operating point. The local linearisation \eqref{eq:sch_Jloc} is unchanged since $\partial[-\chi(v-v_{\mathrm{ref}})]/\partial v=-\chi$.
\section{NOS-Based Scheduler}
\label{sec:nos}
We model each bearer \(i\) is represented by a NOS two-state element \((v_i,u_i)\) \cite{bilal2025NOS}.  
The state excitability \(v_i\) acts as a bounded proxy for the queue backlog, while state \(u_i\) provides a recovery credit that discourages repeated grants. 
These states evolve continuously under the joint influence of arrivals, service leakage, and delayed neighbour activity. 
Parameters \(a>0\) and \(b>0\) set the characteristic recovery and feedback time scales.
The NOS scheduler forms the central link between the queue dynamics of Section~\ref{sec:system} and the slot-level resource allocation mechanism. Figure~\ref{fig:nos_overview_2col} summarises the complete flow. Compound–Poisson arrivals from multiple UEs update per-slot queues that feed the NOS kernel \((v_i,u_i)\). 
Within the kernel, bounded excitability, recovery, and a stabilising leak 
\( -\chi\,(v_i - v_{\mathrm{ref}}) \) interact with delayed neighbour influence \( g\,W\,S(t-\Delta) \) to produce spikes \(S_i(t)\). These spikes are filtered and weighted in a PF sense, aggregated within interference cliques, and mapped to integer PRB grants through floor, filtering, renormalisation, and water-filling steps. 
Near-RT control acts through the effective delay 
\(\Delta_{\mathrm{eff}}\) to adjust the thresholds and the coupling gain \(g\).
\vspace{-4.2mm}
\subsection{State evolution and networking interpretation}
Bounded excitability, leaky service, delayed neighbour influence, and arrival shot noise:
\begin{align}
\begin{split}
\frac{dv_i}{dt} = \underbrace{f_{\mathrm{sat}}(v_i)+\beta v_i}_{\text{bounded growth from backlog}}
                  \;-\; \underbrace{u_i}_{\text{short-term credit}} -\\
                   \underbrace{\chi\,(v_i - v_{\mathrm{ref}})}_{\text{leak for stability}}
                  \;+\; \underbrace{g\!\sum_{j} w_{ij}S_j(t-\Delta)}_{\text{delayed neighbour influence}}
                  \;+\; \eta_i(t), \label{eq:sch_dv_sat}\\
\end{split}
\end{align}

\begin{align}
\begin{split}
\frac{du_i}{dt} &= a\big(b v_i - u_i\big), \label{eq:sch_du}
\end{split}
\end{align}

with
\begin{align}
f_{\mathrm{sat}}(v)&=\frac{\alpha v^2}{1+\kappa v^2}, \quad
f'_{\mathrm{sat}}(v)=\frac{2\alpha v}{(1+\kappa v^2)^2}, \label{eq:fsat_def}\\
\eta_i(t)&=\sum_{n=1}^{N_i(t)} A_{i,n}\,e^{-(t-t_{i,n})/\tau_s}\,\mathbf{1}_{\{t\ge t_{i,n}\}}. \label{eq:shot_noise}
\end{align}
The bounded nonlinearity is globally Lipschitz with
\begin{align}
\begin{split}
\sup_v |f'_{\mathrm{sat}}(v)|=\tfrac{3\sqrt{3}}{8}\,\frac{\alpha}{\sqrt{\kappa}} \doteq L_{\mathrm{sat}},
\\
L \;\le\; \max\!\big\{L_{\mathrm{sat}}+|\beta-\chi|+1,\;ab+a\big\}.
\label{eq:Lip}
\end{split}
\end{align}
\emph{Networking mapping.} \(f_{\mathrm{sat}}\) encodes finite buffers; \(\chi v_i\) is a stabilising leak; \(gW\) captures delayed coordination pressure.

\subsection{Spike generation, fairness, and soft reset}
Spikes request grants. A soft reset models finite grant duration:
\begin{align}
\begin{split}
\text{if } \frac{\mu_i(t)\,v_i(t)}{\bar r_i(t)+\varepsilon} \ge v_{\mathrm{th}}:\;
S_i(t)&=1,\;
v_i \leftarrow c + \big(v_i-c\big)e^{-\nu\,\tau_{\text{rst}}},\\
u_i \leftarrow u_i + d, \label{eq:sch_reset}\;
\text{else }\; S_i(t)=0.\nonumber
\end{split}
\end{align}
Here \(\bar r_i(t)\) is an EWMA of realised rate for PF-like long-term fairness; \(\nu>0\) is the reset rate; \(\varepsilon>0\) prevents division by zero. Choose \(\vartheta\in(0,1]\) in \(\bar r_i\) so that \(\tau_{\bar r}\approx 1/\vartheta\) matches ten to twenty slots.

\subsection{Spike-to-PRB mapping with feasibility and rounding}
Requests are filtered and normalised; interference reduces effective demand through delayed spikes:
\begin{align}
r_i(t) &=(1-\zeta)\,r_i(t-1)+\zeta\,S_i(t), \;\; \zeta\in(0,1], \\
\tilde r_i(t) &= \frac{r_i(t)}{1+\sum_{j} w_{ij}S_j(t-\Delta)}, \label{eq:sch_req}\\
\bar r_i(t) &=(1-\vartheta)\,\bar r_i(t-1)+\vartheta\,\mu_i(t-1)\,x_i(t-1), \;\; \vartheta\in(0,1], \label{eq:sch_rate_ewma}\\
w_i(t)&=\frac{\tilde r_i(t)\,\mu_i(t)}{\bar r_i(t)+\varepsilon}. \label{eq:sch_pfweight}
\end{align}
Let \(\{\mathcal{C}_m\}\) be interference cliques with per-slot PRB budget fractions \(\gamma_{\mathcal{C}_m}\in(0,1]\). For each \(\mathcal{C}_m\),
\begin{align}
x_i^{\text{cont}}(t)=
\begin{cases}
\displaystyle \min\!\Big\{1,\,\gamma_{\mathcal{C}_m}\frac{w_i(t)}{\sum_{j\in\mathcal{C}_m} w_j(t)+\varepsilon}\Big\},\;i\in\mathcal{C}_m,\\[2ex]
0, \;\;\;\;\;\;\;\;i\notin\mathcal{C}_m.
\end{cases} \label{eq:sch_x}
\end{align}
\textit{Integer PRBs.} Within each clique: floor to whole PRBs, drop UEs that fail an MCS minimum, renormalise over survivors to respect \(\gamma_{\mathcal{C}_m}\), then water-fill remaining PRBs to the largest residuals.

\paragraph*{DU slotting and numerical stability}
Use slot length \(h\). Discretise the linear part via an explicit one-step method:
\begin{equation}
\begin{bmatrix}v\\ u\end{bmatrix}_{t+1}
\approx \Big(I + h\,J_{\mathrm{loc}}\Big)\begin{bmatrix}v\\ u\end{bmatrix}_{t}
+ h\,k\,E\,S_{t-\Delta} + \text{nonlinear terms}.
\end{equation}
A conservative bound that avoids numerical artefacts is
\begin{equation}
h \;<\; \min\!\left\{\frac{1}{L},\;\frac{2}{\big\|J_{\mathrm{loc}}+kE\big\|_2+\epsilon}\right\},
\qquad
k=g\,H\,\rho(W),
\label{eq:stepbound}
\end{equation}
where \(H=\big.\tfrac{\partial S}{\partial v}\big|_{v^*}\) is the local spike slope. When the zero-delay crossing has frequency \(\omega_0>0\) (defined in \eqref{eq:omega0}), choose \(h\) so that \(h\,\omega_0\le 1\).

\section{Stability Analysis}
\label{sec:stability}
We study subthreshold equilibria of \eqref{eq:sch_dv_sat}–\eqref{eq:sch_du}. Absorb the mean of \(\eta_i\) into the operating point. Treat the soft reset as a smooth pullback near threshold. With input delay \(\Delta>0\), the Markov state is augmented by the finite delay line.

\subsection{Local linearisation and topology}
Let \((v^*,u^*)\) be a subthreshold equilibrium and set \(\bar d=f'_{\mathrm{sat}}(v^*)+\beta-\chi\).
\begin{align}
J_{\mathrm{loc}} =
\begin{bmatrix}
\bar d & -1\\
ab & -a
\end{bmatrix},
\qquad
E =
\begin{bmatrix}
1 & 0\\
0 & 0
\end{bmatrix}. \label{eq:sch_Jloc}
\end{align}
Linearising the spike map gives \(S(t)\approx H\,v(t)\) near threshold, hence the modal subsystems
\[
\dot{\tilde z}_\ell(t)
=
J_{\mathrm{loc}}\tilde z_\ell(t)
+
k_\ell\,E\,\tilde z_\ell(t-\Delta),
\qquad
k_\ell=g\,H\,\lambda_\ell .
\]

With zero delay, the closed-loop matrix is \(A_0=J_{\mathrm{loc}}+kE\), \(k=g\,H\,\rho(W)\). Routh–Hurwitz yields
\begin{align}
\operatorname{tr}(A_0)&=\bar d-a+k<0 \Rightarrow k<a-\bar d,\\
\det(A_0)&=a\,(b-\bar d - k)>0 \Rightarrow k<b-\bar d,
\end{align}
hence
\begin{equation}
k_\star(0,\bar d,a,b)=\min\{a-\bar d,\; b-\bar d\}.
\label{eq:sch_kstar_nodelay}
\end{equation}

\subsection{Delay-aware threshold and spectral margin}
With an input delay \(\Delta\) on the \(v\) channel, the boundary condition at \(s=\mathrm{j}\omega\) is
\begin{equation}
\det\!\big(sI_2 - J_{\mathrm{loc}} - k\,e^{-\mathrm{j}\omega\Delta}\,E\big)=0.
\label{eq:delayCE}
\end{equation}
This yields two real equations in \((\omega,k)\). The smallest \(k\) occurs at the first crossing frequency \(\omega_\star(\Delta)\). One has \(k_\star(\Delta)\le k_\star(0)\) with equality only at \(\Delta=0\).

\paragraph{Perron–mode proxy and global margin.}
Let \(W\in\mathbb{R}^{N\times N}\) be entrywise non-negative and diagonalizable with eigenpairs \((\lambda_\ell,v_\ell)\). A sufficient test, exact for symmetric \(W\) and accurate for mildly non-normal \(W\), is
\begin{equation}
\delta \doteq k_\star(\Delta) - g\,H\,\rho(W) > 0,
\;
k_\star(\Delta)\equiv \max_\ell k_{\star,\ell}(\Delta).
\label{eq:perron_margin}
\end{equation}

\begin{corollary}[Delay envelope used in numerics]
\label{cor:envelope}
For \(\Delta=0\), \(k_\star(0)=\min\{a-\bar d,\; b-\bar d\}\) from \eqref{eq:sch_kstar_nodelay}. The zero-delay crossing frequency is
\begin{equation}
\omega_0=
\begin{cases}
\sqrt{a(b-a)}, & a\le b,\\[3pt]
0, & b<a.
\end{cases}
\label{eq:omega0}
\end{equation}
For small \(\Delta\) and \(a\le b\), a first–order Padé approximation yields the conservative bound
\(
k_\star(\Delta)\;\gtrsim\; \frac{k_\star(0)}{1+\tfrac{\Delta}{2}\,\omega_0}.
\)
For implementation we adopt the calibrated monotone envelope
\begin{equation}
\label{eq:kstar_envelope}
k_\star(\Delta)=k_\star(0)\,e^{-\Delta/\tau_{\mathrm{ctrl}}},
\end{equation}
with \(\tau_{\mathrm{ctrl}}\) fitted once at \(\Delta=0\) to match the Padé slope.
\end{corollary}

\begin{definition}[Effective margin used in tables and plots]
\label{def:delta_eff}
Split delays into signalling \(\Delta_{\mathrm{sig}}\), observation \(\Delta_{\mathrm{obs}}^{(s)}\), and DU compute \(\Delta_{\mathrm{DU}}^{(s)}\). For \(s\in\{\mathrm{NOS},\mathrm{PF},\mathrm{BP}\}\),
\(
\Delta_{\mathrm{eff}}^{(s)}=\Delta_{\mathrm{sig}}+\Delta_{\mathrm{obs}}^{(s)}+\Delta_{\mathrm{DU}}^{(s)},
\qquad
\Delta_{\mathrm{obs}}^{(\mathrm{BP})}=\phi_{\mathrm{obs}}\,\Delta_{\mathrm{sig}}.
\)
With \(k_\star(\cdot)\) from \eqref{eq:kstar_envelope},
\(
\delta^{(s)} \;=\; k_\star\!\big(\Delta_{\mathrm{eff}}^{(s)}\big)
- \mathbf{1}_{\{s=\mathrm{NOS}\}}\,g\,H\,\rho(W).
\)
For PF and BP this \(\delta^{(s)}\) is a headroom proxy induced by timing, not a closed-loop spectral margin.
\end{definition}

\begin{corollary}[Tail proxy consistent with Theorem~\ref{thm:geo}]
\label{cor:tails}
Let \(\gamma_{\mathrm{eff}}^{(s)}\) be the effective utilisation cap and \(\mathrm{AUC}^{(s)}\) the achieved area under the utilisation–load curve over \([0,\gamma_{\mathrm{eff}}^{(s)}]\), normalised by \(\gamma_{\mathrm{eff}}^{(s)}\). Set \(\bar x^{(s)}=\min\!\big\{1,\mathrm{AUC}^{(s)}/(\gamma_{\mathrm{eff}}^{(s)}+\varepsilon)\big\}\) and fix a service floor \(\mu_{\min}>0\) as the 5th percentile of PRB rate. For a constant \(\kappa_\theta>0\) chosen to match the small-\(\tau\) slope of empirical CCDFs at \(\Delta=0\), define
\[
\theta^{(s)}=\kappa_\theta\,\delta^{(s)}\big(\mu_{\min}\,\bar x^{(s)}\big)^2.
\]
Then the numerical tail bounds used in the figures take the form
\(
\Pr\{D>\tau\}\;\lesssim\;\exp\!\big(-\theta^{(s)}\tau^2\big),
\;
\Pr\{q\ge x\}\;\lesssim\;\exp\!\big(-\theta^{(s)}x^2/\varsigma^2\big),
\)
which matches the structure of Theorem~\ref{thm:geo} when \(\delta^{(s)}>0\).

\end{corollary}

\begin{remark}[Non-normal topology]
If \(W\) is markedly non-normal, replace \(\rho(W)\) in \eqref{eq:perron_margin} by a conservative surrogate such as \(\|W\|_2\) or a small-\(\epsilon\) pseudospectral radius. The code is unchanged after this swap.
\end{remark}

\subsection{Early-warning signal}
Let \(\phi_i(t)\) be the phase of a Hilbert transform of a low-pass filtered spike train \(r_i(t)\) (order 2, cut-off at a few slots). Define
\begin{equation}
R(t)=\frac{1}{N}\Big|\sum_{i=1}^N e^{\mathrm{i}\phi_i(t)}\Big|.
\label{eq:sch_R}
\end{equation}
As a practical alternative, use the principal component ratio of the spike covariance over a rolling window; it rises with synchrony and avoids phase extraction. Decision rule: declare an overload warning when \(R(t)\) exceeds the \(95\)th percentile of its light-load distribution for \(T\) consecutive slots.

\subsection{Mean backlog and tail bounds}
\begin{assumption}
\label{as:noise}
Arrivals have finite exponential moments; \(f_{\mathrm{sat}}\) and the soft reset render the drift globally Lipschitz; the mean offered load lies within capacity.
\end{assumption}

\begin{theorem}[Geometric ergodicity under a spectral margin]
\label{thm:geo}
Under Assumption~\ref{as:noise}, if \(\delta>0\) then the NOS-controlled Markov process for the augmented state consisting of \((v,u)\) and the finite delay line is geometrically ergodic. There exist \(C_0,C_1>0\) such that
\begin{align}
\sum_{i=1}^N \mathbb{E}[v_i^2] \le \frac{C_1}{2C_0\,\delta}, 
\qquad
\mathbb{E}[q_i] \le \varsigma\Big(\frac{C_1}{2C_0\,\delta}\Big)^{1/2}.
\label{eq:sch_meanq}
\end{align}
Moreover, with the global Lipschitz constant \(L\) from \eqref{eq:Lip} there exists \(\theta_\star \ge (C_0/L)\,\delta\) so that for any \(0<\theta<\theta_\star\),
\begin{align}
\Pr\{q_i \ge x\} &\le K_i(\theta)\,\exp\!\Big(-\theta\,x^2/\varsigma^2\Big), \label{eq:sch_qtail}\\
\Pr\{D_i>\tau\} &\le \tilde K_i(\theta)\,\exp\!\Big(-\theta\,(\mu_i^{\min}\bar x_i)^2\,\tau^2/\varsigma^2\Big), \label{eq:sch_tails}
\end{align}
with \(\mu_i^{\min}>0\) and \(\bar x_i=\mathbb{E}[x_i]>0\). If arrivals are subexponential, tails inherit that heaviness beyond a crossover scale; the geometric bound still applies near the typical operating regime.
\end{theorem}

\paragraph*{Design notes (networking context)}
Neighbour pressure may enter the state path (as in \eqref{eq:sch_dv_sat}) or only the grant path via \(\tilde r_i\). Near threshold the small-signal gain from neighbour spikes to \(v\) is \(gH\) up to topology, so the Perron-mode margin \(k=gH\rho(W)\) applies in both cases. The single knob \(\delta=k_\star-k\) aggregates topology \(\rho(W)\), delay \(\Delta\), and controller strength \(g\).

\subsection{Complexity and DU realisation}
\label{subsec:complexity-du}
\textbf{Per-slot work.}
Forming neighbour pressure $\sum_j w_{ij}S_j$ costs $O(|\mathcal{E}|)$ for edge set $\mathcal{E}$ of the interference graph. Clique normalisation in \eqref{eq:sch_x} is $O(\sum_m |\mathcal{C}_m|)$ per slot. The PRB rounding step (floor, drop, renormalise, water-fill) is $O(|\mathcal{C}_m|\log |\mathcal{C}_m|)$ per clique using a heap; linear-time water-fill is possible for fixed PRB budgets.

\textbf{Slotting and numerical stability.}
With slot length $h$, integrate the linear part using an explicit one-step method and add the nonlinear terms bounded by \(L\) in \eqref{eq:Lip}. A conservative bound that avoids numerical artefacts is
\begin{equation}
h \;<\; \min\!\left\{\frac{1}{L},\;\frac{2}{\big\|J_{\mathrm{loc}}+kE\big\|_2+\epsilon}\right\},
\quad
k=g\,H\,\rho(W).
\label{eq:du-stepbound}
\end{equation}
When \(\omega_0>0\) in \eqref{eq:omega0}, keep \(h\,\omega_0\le 1\).

\textbf{Grant feasibility on integer PRBs.}
Within each interference clique $\mathcal{C}_m$:
(i) compute continuous $x_i^{\text{cont}}$ by \eqref{eq:sch_x},
(ii) map to whole PRBs by flooring,
(iii) drop users whose transport block would fall below a target size and renormalise over survivors, and
(iv) water-fill remaining PRBs to the largest residuals. This preserves $\sum_{i\in\mathcal{C}_m} x_i\le \gamma_{\mathcal{C}_m}\le 1$ and keeps the spike-to-grant characteristics.

\textbf{Where the control lives.}
Neighbour pressure may enter the state path (as in \eqref{eq:sch_dv_sat}) or the grant path only (via the denominator in $\tilde r_i$). Near threshold both induce the same small-signal gain up to the factor \(H\) absorbed in \(k=gH\rho(W)\). Near-RT xApps modulate \(g\) and thresholds; the DU executes the per-slot loop with the chosen \(g\).

Table~\ref{tab:init} gives safe defaults values for DU deployment. Set the recovery horizon by $a$, tune $b$ to match PF reactivity, choose $(\alpha,\kappa)$ so that the saturating slope $L_{\mathrm{sat}}=\tfrac{3\sqrt{3}}{8}\alpha/\sqrt{\kappa}$ stays below a configured bound, and pick $(c,d,\tau_{\text{rst}})$ to model a short cool-down after a grant. The EWMA $(\zeta,\vartheta)$ control the request and rate memories (typical time constants 10--20 slots). Finally, select a policy headroom $\delta$as a fraction of $k_\star(\Delta)$ and set $g=(k_\star-\delta)/(H\,\rho(W))$. Units: if $q=\varsigma v$, report $\varsigma$ once.
\vspace{-2.5mm}
\begin{table}[h]
\centering
\caption{Parameter initialisation and nominal ranges for DU realisation}
\label{tab:init}
\begin{tabular}{@{}ll@{}}
\toprule
Parameter & Choice / range and rationale \\
\midrule
$a$ & $1/\tau_{\mathrm{fair}}$, $\tau_{\mathrm{fair}}=50$–$100$\,ms (credit decay) \\
$b$ & $0.5$–$1.5$ (match PF short-term responsiveness) \\
$\chi$ & match $\overline{\mu}\,\overline{x}$ at light load (service leak) \\
$\alpha,\kappa$ & set $L_{\mathrm{sat}}=\tfrac{3\sqrt{3}}{8}\alpha/\sqrt{\kappa} \le L_{\max}$ \\
$c$ & near light-load $v$ (post-spike baseline) \\
$d$ & small positive (cool-down increment in $u$) \\
$\tau_{\text{rst}}$ & 2–4 slots (finite grant duration) \\
$\zeta$ & $0.1$–$0.3$ (request LPF) \\
$\vartheta$ & $\tau_{\bar r}\approx 10$–20 slots (rate-EWMA gain) \\
$\nu$ & reset rate in $e^{-\nu\tau_{\text{rst}}}$ \\
$\delta$ & $0.1$–$0.3$ of $k_\star(\Delta)$ (policy headroom) \\
$\gamma_{\mathrm{ref}}$ & tail normalisation: $\gamma_{\mathrm{eff}}$ (default) or $\gamma_c^{\mathrm{nom}}$ (DU-neutral) \\
\bottomrule
\end{tabular}
\end{table}
\vspace{-2.5mm}
\section{Experimental Setup and Results}
\label{sec:exp}

In this section we first describe the experimantal setup, and later discuss the results. We use a $\delta$-proxy numerical study consistent with Sections~\ref{sec:system}–\ref{sec:stability}. Local parameters are \(a=1.0\), \(b=0.9\), and \(\bar d=0.30\), hence \(k_\star(0)=\min\{a-\bar d,b-\bar d\}=0.60\). Delay attenuation follows the envelope \(k_\star(\Delta)=k_\star(0)\exp(-\Delta/\tau_{\mathrm{ctrl}})\) with \(\tau_{\mathrm{ctrl}}=10\)~ms, as in \eqref{eq:kstar_envelope}. Interference uses three symmetric graphs with uniform weight \(w=0.60\): pair2, line4, and ring8. We compute \(\rho(W)\) from the eigenvalues of \(W\); for reproducibility, \(\rho(\text{pair2})=0.60\), \(\rho(\text{line4})\approx 0.971\), and \(\rho(\text{ring8})=1.20\). The sweep is \(\Delta\in\{5,12,20\}\)~ms.

DU processing follows Section~\ref{subsec:complexity-du} with a \(120~\mu\)s per-slot budget and \(U_c=8\) users per cell. Per-cell costs are NOS \((22+1.2\,U_c+0.06\,E/C)\,\mu\text{s}\), PF \((28+1.8\,U_c)\,\mu\text{s}\), and BP \((38+2.7\,U_c+0.14\,E/C)\,\mu\text{s}\). Compute headroom scales the radio cap from \(\gamma_c^{\mathrm{nom}}=0.95\) to \(\gamma_{\mathrm{eff}}\); any overflow adds a capped DU spillover \(\Delta_{\mathrm{DU}}\le 1\)~ms. BP carries an observation lag \(0.15\,\Delta\) (Definition~\ref{def:delta_eff}). Effective delays are \(\Delta_{\mathrm{eff}}^{(\mathrm{NOS})}=\Delta+\Delta_{\mathrm{DU}}^{(\mathrm{NOS})}\), \(\Delta_{\mathrm{eff}}^{(\mathrm{PF})}=\Delta+\Delta_{\mathrm{DU}}^{(\mathrm{PF})}\), and \(\Delta_{\mathrm{eff}}^{(\mathrm{BP})}=(1+0.15)\Delta+\Delta_{\mathrm{DU}}^{(\mathrm{BP})}\).

A single coupling gain \(g\) is fixed once at the worst case (largest \(\rho(W)\), \(\Delta=20\)~ms) by tail-parity calibration: NOS \(p_{99.9}\) is matched to PF \(p_{99.9}\) at that design point (the \texttt{match\_p999\_floor} rule in code). The same \(g\) is then used for all topologies and delays. In numerics the spike slope is absorbed, so \(H=1\). For scheduler \(s\in\{\mathrm{NOS},\mathrm{PF},\mathrm{BP}\}\) the working margin follows Definition~\ref{def:delta_eff} with \(\delta^{(s)}=k_\star(\Delta_{\mathrm{eff}}^{(s)})-\mathbf{1}_{\{s=\mathrm{NOS}\}}\,g\,\rho(W)\), and we take \(\eta_{\mathrm{PF}}=\eta_{\mathrm{BP}}=1\).
\textcolor{black}{ To avoid confusion with classifier ROC-AUC, we report a \emph{normalised utilisation headroom} $\mathrm{AUC}_{\mathrm{util}}$ over the load range, defined as the area under the utilisation--load curve divided by the effective cap. In code this is implemented by the saturating mapping below, and the plotted quantity is the dimensionless ratio \(\mathrm{AUC}_{\mathrm{util}}/\gamma_{\mathrm{eff}}\in[0,1]\). Values in the \(0.4\)–\(0.7\) range therefore indicate moderate headroom under the chosen margins and delays, not poor forecasting accuracy.}
\textcolor{black}{Utilisation is mapped from \(\delta\) by the saturating law implemented in the code,}
\(
\textcolor{black}{\mathrm{AUC}_{\mathrm{util}}=\gamma_{\mathrm{eff}}\;\frac{\kappa_u\,\delta}{1+\kappa_u\,\delta},}
\)
\textcolor{black}{with \(\kappa_u=25\) and a clamp at \(0.05\,\gamma_{\mathrm{eff}}\) when \(\delta\le 0\).}

Tail quantities use Corollary~\ref{cor:tails} with \(\mu_{\min}=0.12\) pkts/slot and \(\bar x=\min\{1,\mathrm{AUC}_{\mathrm{util}}/\gamma\}\), yielding
\(
p_{99.9}=\frac{1}{\mu_{\min}\bar x}\sqrt{\frac{\log 1000}{\kappa_\theta\,\delta}}
\quad\text{and}\quad
\mathrm{MaxQ}\approx \sqrt{\frac{\log 1000}{\kappa_\theta\,\delta}},
\)
with \(\kappa_\theta=4.0\). Two normalisations are reported: the nominal-cap mode \(\bar x=\mathrm{AUC}_{\mathrm{util}}/\gamma_c^{\mathrm{nom}}\) used for default tables and averaged curves, and the effective-cap mode \(\bar x=\mathrm{AUC}_{\mathrm{util}}/\gamma_{\mathrm{eff}}\) shown in figures as a sensitivity check.




\begin{figure*}[t]
\centering
\includegraphics[width=\textwidth]{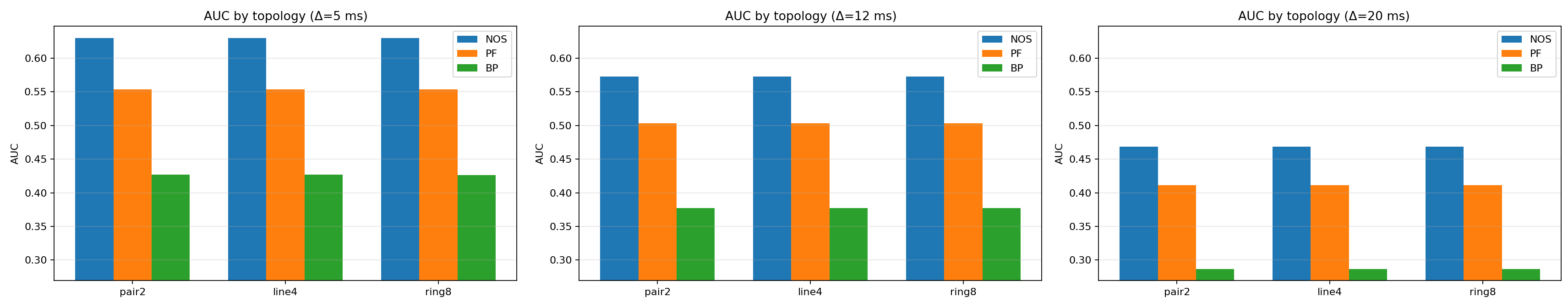}
\caption{Utilisation headroom AUC across pair2, line4, and ring8 at \(\Delta\in\{5,12,20\}\) ms. Nominal-cap normalisation; setup as in Sec.~\ref{sec:exp}.}
\label{fig:auc_bars}
\end{figure*}

As shown in Fig.~\ref{fig:auc_bars}, NOS attains the highest AUC for every topology and delay. As \(\Delta\) increases from 5 to 20 ms, all schedulers lose headroom through the envelope \(k_\star(\Delta)\), so AUC falls in parallel; the gap between NOS and PF narrows at the largest delay. BP remains lowest because it carries both the observation penalty and the higher DU cost, which reduce \(\gamma_{\mathrm{eff}}\). Differences across pair2, line4, and ring8 are small, consistent with calibrating a single \(g\) at the worst spectral radius and reusing it; delay dominates the mapping from margin to AUC.

\begin{figure*}[t]
\centering
\includegraphics[width=.9\textwidth]{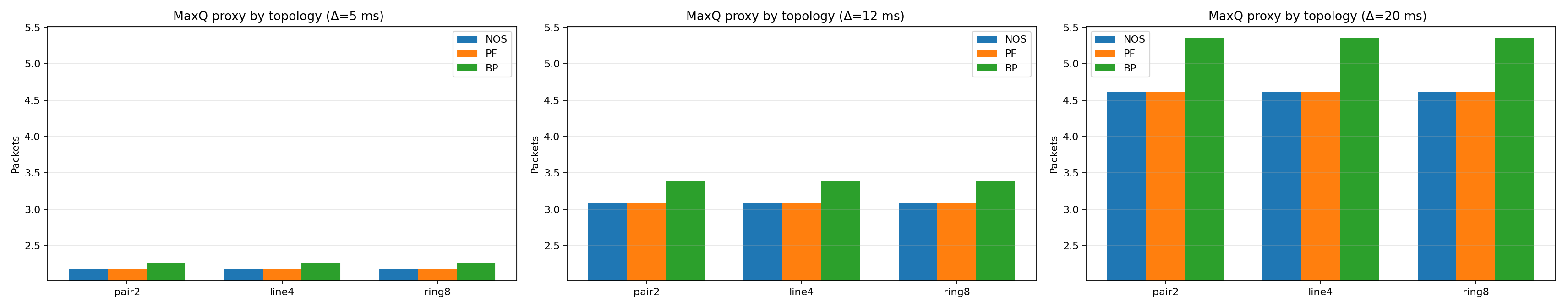}
\caption{MaxQ proxy across pair2, line4, and ring8 at \(\Delta\in\{5,12,20\}\) ms, computed from \(\delta\) as in Corollary~\ref{cor:tails}. Nominal-cap normalisation; setup as in Sec.~\ref{sec:exp}.}
\label{fig:maxq_bars}
\end{figure*}

Figure~\ref{fig:maxq_bars} shows a monotone increase of the MaxQ proxy with \(\Delta\), reflecting shrinkage of the margin under \(k_\star(\Delta)\). BP yields the largest MaxQ at \(\Delta\ge 12\) ms due to the combined observation penalty and higher DU cost, which reduce \(\gamma_{\mathrm{eff}}\) and enlarge delays. NOS and PF are essentially coincident because \(g\) was set by tail parity at the worst case and, in nominal-cap mode, the proxy depends only on \(\delta\); both therefore track each other across the sweep. Topology effects are small once a single \(g\) is fixed at the worst spectral radius.

\begin{figure*}[t]
\centering
\includegraphics[width=.9\textwidth]{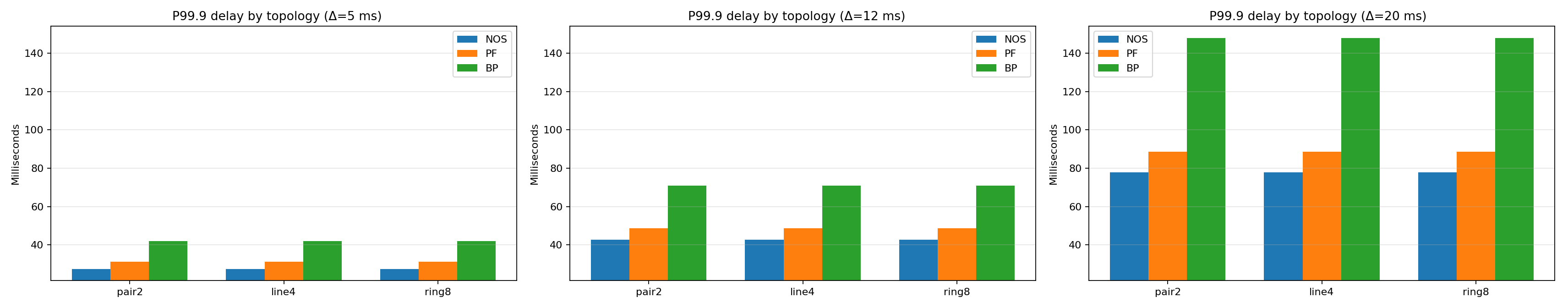}
\caption{P99.9 delay across pair2, line4, and ring8 at \(\Delta\in\{5,12,20\}\) ms. Nominal-cap normalisation; setup as in Sec.~\ref{sec:exp}.}
\label{fig:p999_bars}
\end{figure*}

Figure~\ref{fig:p999_bars} shows a steady increase of the P99.9 delay with \(\Delta\), the direct effect of the envelope \(k_\star(\Delta)\) reducing the spectral margin. BP is consistently the largest owing to its observation lag and higher DU cost, both of which lower \(\gamma_{\mathrm{eff}}\) and widen the tail. NOS remains the smallest across the sweep; at \(\Delta=20\) ms it lies close to PF, as expected from the tail-parity calibration used to set \(g\). Differences across pair2, line4, and ring8 are modest once a single \(g\) is fixed at the worst spectral radius, so control latency is the dominant factor.

\begin{figure*}[t]
\centering
\includegraphics[width=.9\textwidth]{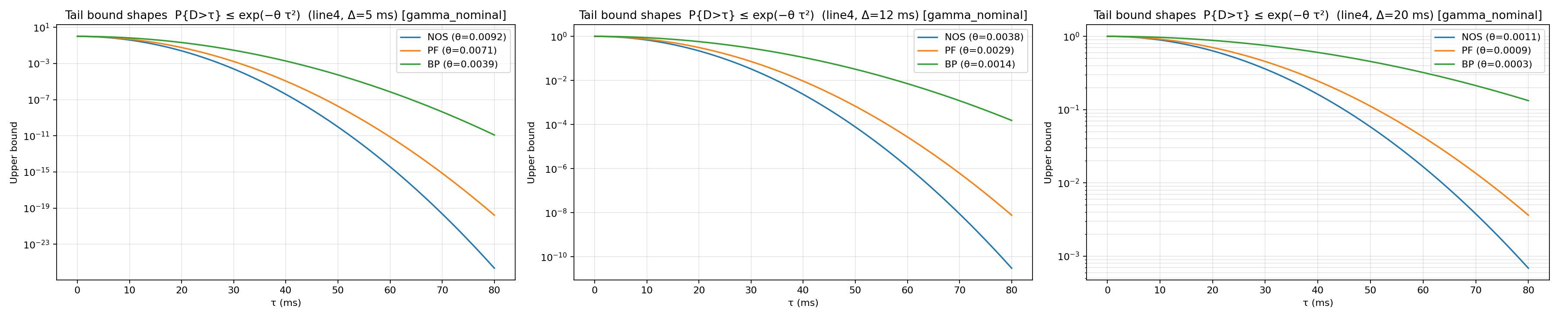}
\caption{Analytic delay tail bounds for \texttt{line4} on a log scale. Bounds follow \(\Pr\{D>\tau\}\le \exp(-\theta\tau^2)\) with \(\theta\) from Corollary~\ref{cor:tails}. Parameters and normalisation as in Sec.~\ref{sec:exp}.}
\label{fig:tail_bounds}
\end{figure*}

Figure~\ref{fig:tail_bounds} shows progressive thinning of the delay tail as the margin increases and, conversely, thickening as \(\Delta\) grows from 5 to 20 ms. NOS has the steepest decay (largest \(\theta\)) at each delay, PF is next, and BP is the loosest bound owing to the observation penalty and higher DU cost that reduce \(\gamma_{\mathrm{eff}}\). At \(\Delta=20\) ms the NOS and PF curves are close, consistent with the calibration used to set \(g\). The absolute values of \(\theta\) summarise the ordering already seen in the AUC and P99.9 plots; the log scale exposes the sub-Gaussian form.

\paragraph*{Results summary.}
Across \(\Delta\in\{5,12,20\}\) ms, NOS achieves the highest utilisation and the tightest delay tails. PF is competitive but consistently below NOS, while BP lags due to observation lag and higher DU cost. With a single \(g\) fixed at the worst spectral radius, topology has only a mild effect; control latency through \(k_\star(\Delta)\) is dominant. These trends are consistent across AUC (Fig.~\ref{fig:auc_bars}), MaxQ (Fig.~\ref{fig:maxq_bars}), P99.9 (Fig.~\ref{fig:p999_bars}), and the analytic tail shapes (Fig.~\ref{fig:tail_bounds}). The evaluation is deterministic and fully specified by the setup, so tables and figures regenerate from the released code and CSV outputs.
\section{Conclusion}
This work provides an explainable scheduler that links a bounded two-state spiking kernel to a PF-compatible spike-to-PRB head and makes delay explicit through the margin \(\delta=k_\star(\Delta)-gH\rho(W)\). We proved geometric ergodicity for \(\delta>0\) and derived sub-Gaussian backlog and delay bounds with exponents proportional to \(\delta\). A calibrated, deterministic study that folds DU compute limits and observation lag into \(\Delta_{\mathrm{eff}}\) shows consistent gains: with one gain fixed at the worst spectral radius, NOS delivers higher utilisation and smaller P99.9 delay than PF and delayed backpressure across pair2, line4, and ring8 over 5–20 ms.

\bibliographystyle{IEEEtran}
\bibliography{References}

\begin{thebibliography}{10}
\providecommand{\url}[1]{#1}
\csname url@samestyle\endcsname
\providecommand{\newblock}{\relax}
\providecommand{\bibinfo}[2]{#2}
\providecommand{\BIBentrySTDinterwordspacing}{\spaceskip=0pt\relax}
\providecommand{\BIBentryALTinterwordstretchfactor}{4}
\providecommand{\BIBentryALTinterwordspacing}{\spaceskip=\fontdimen2\font plus
\BIBentryALTinterwordstretchfactor\fontdimen3\font minus \fontdimen4\font\relax}
\providecommand{\BIBforeignlanguage}[2]{{%
\expandafter\ifx\csname l@#1\endcsname\relax
\typeout{** WARNING: IEEEtran.bst: No hyphenation pattern has been}%
\typeout{** loaded for the language `#1'. Using the pattern for}%
\typeout{** the default language instead.}%
\else
\language=\csname l@#1\endcsname
\fi
#2}}
\providecommand{\BIBdecl}{\relax}
\BIBdecl

\bibitem{TassiulasEphremides1992}
L.~Tassiulas and A.~Ephremides, ``Stability properties of constrained queueing systems and scheduling policies for maximum throughput in multihop radio networks,'' \emph{IEEE Transactions on Automatic Control}, vol.~37, no.~12, pp. 1936--1948, 1992.

\bibitem{Bonald2003MobiCom}
T.~Bonald and A.~Prouti\`ere, ``Wireless downlink data channels: user performance and cell dimensioning,'' in \emph{Proc. ACM MobiCom}, 2003, pp. 339--352.

\bibitem{Capozzi2013PFsurvey}
F.~Capozzi, G.~Piro, L.~A. Grieco, G.~Boggia, and P.~Camarda, ``Downlink packet scheduling in {LTE} cellular networks: Key design issues and a survey,'' \emph{IEEE Communications Surveys \& Tutorials}, vol.~15, no.~2, pp. 678--700, 2013.

\bibitem{Polese2023ORAN}
M.~Polese, L.~Bonati, S.~D’Oro, S.~Basagni, and T.~Melodia, ``Understanding o-ran: Architecture, interfaces, algorithms, security, and research challenges,'' \emph{IEEE Communications Surveys \& Tutorials}, vol.~25, no.~2, pp. 1376--1411, 2023.

\bibitem{Bonati2023OpenRANGym}
L.~Bonati, M.~Polese, S.~D'Oro, S.~Basagni, and T.~Melodia, ``Openran gym: Ai/ml development, data collection, and testing for o-ran on {PAWR} platforms,'' \emph{Computer Networks}, vol. 220, p. 109502, 2023.

\bibitem{Wu2024R3}
Y.~Wu, Y.~Shi, Y.~T. Hou, W.~Lou, J.~H. Reed, and L.~A. {DaSilva}, ``R\textsuperscript{3}: A real-time robust mu-mimo scheduler for o-ran,'' \emph{IEEE Transactions on Wireless Communications}, vol.~23, no.~11, pp. 17\,727--17\,743, Nov. 2024.

\bibitem{Longhi2025TailORAN}
\BIBentryALTinterwordspacing
N.~Longhi, S.~D'Oro, L.~Bonati, M.~Polese, R.~Verdone, and T.~Melodia, ``Tailo-ran: O-ran control on scheduler parameters to tailor ran performance,'' in \emph{Proc. IEEE GLOBECOM}, 2025, to appear;. [Online]. Available: \url{https://arxiv.org/abs/2508.12112}
\BIBentrySTDinterwordspacing

\bibitem{bilal2025NOS}
\BIBentryALTinterwordspacing
M.~Bilal, ``Network-optimised spiking neural network for event-driven networking,'' 2025. [Online]. Available: \url{https://arxiv.org/abs/2509.23516}
\BIBentrySTDinterwordspacing

\bibitem{Izhikevich2004}
E.~M. Izhikevich, ``Which model to use for cortical spiking neurons?'' \emph{IEEE Transactions on Neural Networks}, vol.~15, no.~5, pp. 1063--1070, 2004.

\bibitem{Liu2024EEDSNN}
Y.~Liu, Z.~Qin, and G.~Y. Li, ``Energy-efficient distributed spiking neural network for wireless edge intelligence,'' \emph{IEEE Transactions on Wireless Communications}, vol.~23, no.~9, pp. 10\,683--10\,697, Sep. 2024.

\bibitem{Liu2001Opportunistic}
X.~Liu, E.~K.~P. Chong, and N.~B. Shroff, ``Opportunistic transmission scheduling with resource-sharing constraints in wireless networks,'' \emph{IEEE Journal on Selected Areas in Communications}, vol.~19, no.~10, 2001.

\end{thebibliography}

\end{document}